\documentclass[12pt]{article}
\usepackage{epsfig,amsfonts,amssymb,amsmath}
\usepackage[margin=2.5cm]{geometry}
\usepackage[pdftex, breaklinks=true, linktocpage,
colorlinks=true, urlcolor=blue, linkcolor=blue, citecolor=red
]{hyperref}
\usepackage[all]{hypcap}
\usepackage{cite}
\usepackage{hyperref}
\usepackage{framed}
\usepackage{breqn}
\usepackage{flexisym}
\usepackage[english]{babel}

\usepackage{ifpdf}
\usepackage[utf8]{inputenc}
\usepackage{longtable}
\usepackage{young}
\usepackage[makeroom]{cancel}
\usepackage[normalem]{ulem}
\usepackage{verbatim}
\usepackage{fancyhdr}
\usepackage[all]{hypcap}
\usepackage{bm, bbm}
\usepackage[usenames, dvipsnames]{xcolor}

\usepackage{eurosym, xspace}
\usepackage[nottoc]{tocbibind}
\usepackage{graphicx, subfig, float}
\usepackage[multiple]{footmisc}
\usepackage{authblk}

\graphicspath{{images/}}
\usepackage[sort&compress, english]{cleveref}

\input epsf.sty
\newcommand{\refb}[1]{(\ref{#1})}
\newcommand{\Tr}{\mathop{\mathrm{Tr}}}
\newcommand{\Res}{\mathop{\mathrm{Res}}}
\newcommand{\sgn}{\mathop{\mathrm{sgn}}}
\newcommand{\nc}{\newcommand}
\nc{\bea}{\begin{eqnarray}}
\nc{\eea}{\end{eqnarray}}
\nc{\be}{\begin{equation}}
\nc{\ee}{\end{equation}}
\nc{\cO}{{\cal O}}
\nc{\cF}{{\cal F}}
\nc{\non}{{\nonumber}}
\nc{\eq}[1]{{(\ref{#1})}}
\nc{\hK}{\widehat{K}}

\begin{document}

\baselineskip 24pt

\begin{center}
{\Large \bf Tensor Models for Black Hole Probes}

\vskip .6cm
\medskip

\vspace*{4.0ex}

\baselineskip=18pt
{\large \rm Nick Halmagyi and Swapnamay Mondal}
\vspace*{4.0ex}

\textit{
\vskip0.2cm
Sorbonne Universit\'es, UPMC Paris 06,  \\ 
UMR 7589, LPTHE, 75005, Paris, France \\
\vskip0.2cm
and \\
\vskip0.2cm
CNRS, UMR 7589, LPTHE, 75005, Paris, France}\\
\vskip0.5cm

\vspace*{1.0ex}
\small swapno@lpthe.jussieu.fr\\ halmagyi@lpthe.jussieu.fr
\end{center}
\vspace*{5.0ex}

\centerline{\bf Abstract} \bigskip
The infrared dynamics of the SYK model, as well as its associated tensor models, exhibit some of the non trivial features expected of a holographic dual of near extremal black holes. These include developing certain symmetries of the near horizon geometry and exhibiting maximal chaos. In this paper we present a generalization of these tensor models to include fields with fewer tensor indices and which can be thought of as describing probes in a black hole background. In large $N$ limit, dynamics of the original model remain unaffected by the probe fields  and the four point functions of the probe fields exhibit maximal chaos, a non trivial feature expected of a black hole probe. Interestingly probe primaries have the same dimensions as primaries of the original fields.
\newpage

\tableofcontents

\newpage

\section{Introduction}  \label{s1}

The study of quantum mechanical models dual to gravitational systems in two dimensions remains a fascinating and difficult arena of research. Quite notably, simple and solvable examples of this duality have proved to be difficult to construct. Somewhat recently however, the fermionic quantum mechanics of SYK model \cite{K2} has been proposed as a system holographically dual to gravity and this has been studied extensively \cite{Maldacena:2016hyu, Maldacena:2016upp, Garcia-Garcia:2017pzl,Sonner:2017hxc, Fu:2016vas, Hunter-Jones:2017raw, Berkooz:2016cvq, Turiaci:2017zwd, Gross:2016kjj, Polchinski:2016xgd, Gross:2017hcz, Gross:2017aos, Das:2017pif}.
A key motivating factor for proposing the SYK model as a holographic dual of a black hole background is the fact that the time out of order four-point correlation functions saturate the so-called {\it maximal chaos bound} \cite{Maldacena:2015waa} which has been shown to hold in the bulk.
Another important feature of the SYK model is that the emergent conformal symmetry is both spontaneously and explicitly broken, which suggests that it is dual to a near AdS2 background.

In an interesting development, it has been shown \cite{Witten:2016iux} that to leading order in the large $N$ expansion, the SYK model (which is disordered, hence not fully quantum mechanical) is identical to the fermionic tensor model of \cite{Gurau:2010ba, Gurau:2011aq}. This has been subsequently generalized in a number of interesting directions \cite{Klebanov:2016xxf, Peng:2016mxj, Krishnan:2016bvg, Choudhury:2017tax, Bulycheva:2017ilt, Peng:2017kro, Yoon:2017nig}.

In this paper we couple the Klebanov-Tarnopolsky model \cite{Klebanov:2016xxf} and Gurau-Witten model\cite{Witten:2016iux} to lower-index fields and interpret the resulting quantum mechanical model as holographically dual to probes in a black hole background. This is in part motivated by  examples where matrix models have been coupled to vector matter, where the latter can be considered as probes in a background described by the former\footnote{See \cite{Iizuka:2008hg}, \cite{Iizuka:2008eb} for an interesting recent example relevant for black hole physics, although this model is not maximally chaotic \cite{Michel:2016kwn}}. The models we study are obtained by adding interactions between the original tensors of  \cite{Witten:2016iux, Klebanov:2016xxf} and tensors of lower rank\footnote{A model which couples rank three and rank one tensor fields has been considered in \cite{Peng:2017kro} but in our model we preserve the interactions purely between the three-index fields ($D$ index fields for color $D$).} .
To leading order in the $1/N$ expansion, the additional  tensors do not affect the physics of original tensors, thus they can rightfully be thought of as probes. Furthermore, we find that all the four point functions exhibit maximal chaos and it is this feature which qualifies these models as toy models for probes in a black hole background. We find that our models have the curious feature that the dimensions of primaries made of out of probe tensors, are identical to those of the original primaries.

The rest of the paper is organized as follows. In \ref{s2}, we give brief introduction to the Klebanov-Tarnopolsky model \cite{Klebanov:2016xxf} and discuss a class of possible modifications, that remain solvable in large $N$ limit and in deep infrared. In \ref{s3}, we consider simplest such model and discuss the propagators, four point functions, primaries and Lyapunov coefficients. We also comment on possible further modifications of this model, that retain the necessary physics. In \ref{s4}, we discuss similar modifications of Gurau-Witten model \cite{Witten:2016iux}. Finally in \ref{s5} we discuss future directions.

\section{Interactions in the D=3 uncolored model} \label{s2}
\subsection{A lightening review of the KT Model}
\begin{figure}[H]
	\begin{center}
		\includegraphics[scale=0.5]{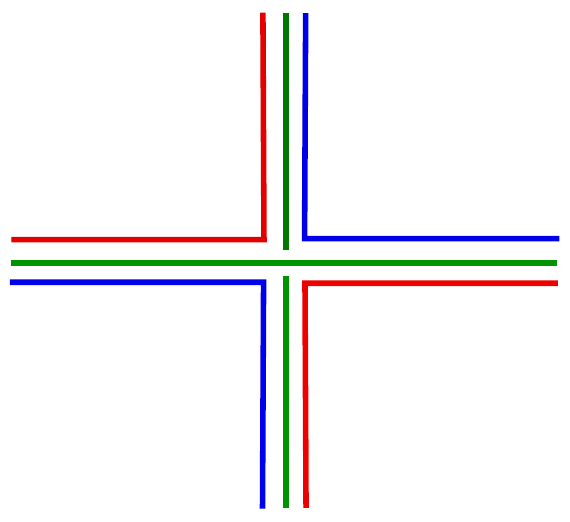}
		\caption{Diagrammatic representation of vertices of KT model: colored lines represent resolved indices. Red, green and blue lines represent respectively first, second and third indices of a the field $\psi_{abc}$.}	\label{klebfig}
	\end{center}
\end{figure}

\begin{figure}[H]
	\begin{center}
		\includegraphics[scale=0.5]{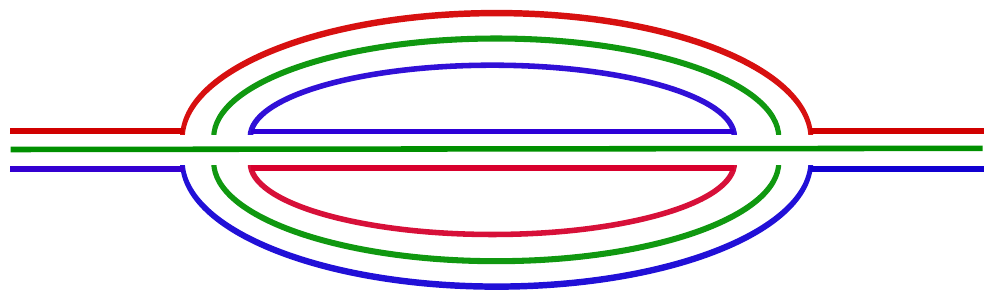}
		\caption{simplest melonic correction to the propagator.}\label{klebprop}
	\end{center}
\end{figure}

The KT model \cite{Klebanov:2016xxf} contains a single real fermionic tensor $\psi_{abc}$ of rank $3$. Each index transforms as a vector under $SO(N)$. To differentiate the three copies of$SO(N)$ we write the first as $SO(N)_1$, the second as $SO(N)_2$ and the third as $SO(N)_3$. The Hamiltonian is taken to be 
\begin{align}
H &= \frac{g_0}{4N^{3/2}} \psi_{a_1b_1c_1} \psi_{a_1b_2c_2} \psi_{a_2b_1c_2} \psi_{a_2b_2c_1} \label{kleb}
\end{align}
whose diagrammatic representation is given in fig \ref{klebfig}.
This model can be obtained by ``uncoloring"\cite{Bonzom:2012hw} the D=3 Gurau-Witten model \cite{Witten:2016iux}. The large $N$ limit \cite{Gurau:2010ba}, \cite{Gurau:2011aq} is defined as taking $N \rightarrow \infty$ while keeping $g_0$ fixed. In this limit it is the melonic grpahs which contribute to leading order in $1/N$ and the simplest correction to the propagator comes from the melonic graph in fig \ref{klebprop}.
A factor of $N^3$, coming from fields propagating in loops, cancels the $1/N^3$ coming from the two vertices, giving an overall factor of $N^0$. Additional melonic corrections to the propagator of the same order are obtained by replacing any of the internal or external propagators in the diagram by this melonic diagram itself. In the large $N$ limit this class of diagrams\footnote{Joining the ends of a propagator gives a vacuum diagram. Thus this class of diagrams also give leading contributions to free energy in large $N$ limit. Joining the ends turns the external lines into internal ones and thus one gets an extra factor of $N^3$. Since the propagators were $\mathcal{O}(N^0)$, this means that free energy scales as $N^3$, which is good since the number of fields scales as $N^3$.} constitute the complete leading corrections to the free propagator. They can be summed up to give the exact propagator in deep IR as follows. Denoting $G(t_1,t_2)$ to be the propagator and $\Sigma(t_1,t_2)$ to be the 1PI two point function to leading order in $1/N$, one has
\begin{align}
\Sigma(t_1,t_2) &= g_0^2 G(t_1,t_2)^3 \, .
\end{align}
By definition
\begin{align}
G(\omega) &= \frac{1}{-i\omega - \Sigma(\omega)} \, .
\end{align}
and in the deep infrared, $i\omega$ can be ignored. In position space, one obtains
\begin{align}
\int dt G(t_1,t) \Sigma(t, t_2) &= -\delta(t_1-t_2) \, .
\end{align}
giving the following Schwinger-Dyson equation
\begin{align}
g_0^2 \int dt G(t_1,t) G(t,t_2)^3 &= -\delta(t_1-t_2) \, , \label{KlebSD}
\end{align}
in deep IR. The result \refb{KlebSD} is invariant under the conformal transformations\footnote{In one dimension conformal group contains all reparameterizations.} $t \rightarrow f(t)$
\begin{align}
G(t_1,t_2) \rightarrow \left| \frac{df(t_1)}{dt_1} \frac{df(t_2)}{dt_2} \right|^{1/4} G(f(t_1), f(t_2)), ~~~~ \Sigma(t_1,t_2) \rightarrow \left| \frac{df(t_1)}{dt_1} \frac{df(t_2)}{dt_2} \right|^{3/4} \Sigma(f(t_1), f(t_2)) \, . \label{conf}
\end{align}
Thus the system develops an emergent conformal symmetry in the deep IR. The solution to \refb{KlebSD} is
\begin{align}
G_c(t) &= \frac{b}{|t|^{1/2}} \sgn(t),~~~~\text{where,}~~b^4  = \frac{1}{4 \pi g_0^2} \, . \label{klebpropa}
\end{align}
which spontaneously breaks the conformal symmetry to $SL(2, \mathbb{R})$.

\begin{figure}[H]
	\begin{center}
		\includegraphics[scale=0.7]{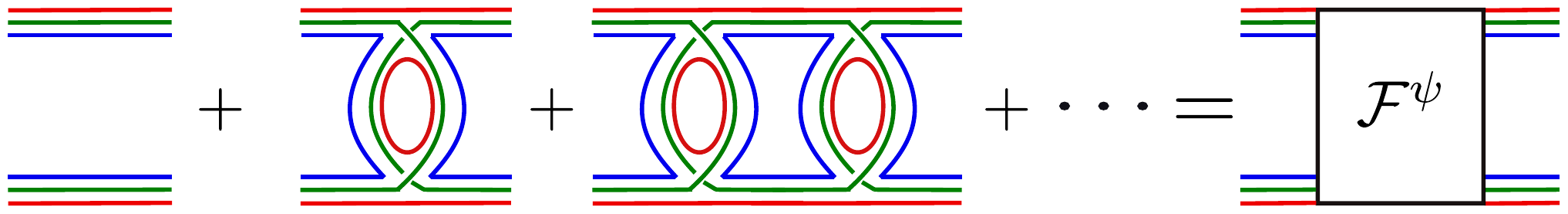}
		\caption{Leading contributions to ``gauge invariant four point function" in large $N$ limit.}	\label{klebladder}
	\end{center}
\end{figure}

Next one considers the ``gauge invariant" four point function, which has the following structure
\begin{align}
\frac{1}{N^6} \sum_{a_1, b_1, c_1, \atop a_2, b_2, c_2} \langle \psi_{a_1b_1c_1}(t_1)  \psi_{a_1b_1c_1}(t_2)  \psi_{a_2 b_2 c_2}(t_3) \psi_{a_2 b_2 c_2}(t_4) \rangle &= G(t_1,t_2) G(t_3,t_4) + \frac{1}{N^3} \mathcal{F}(t_1,t_2;t_3,t_4) \, , \label{psi4defn}
\end{align} 
where $\mathcal{F}$ is given by the sum of ladder diagrams shown in fig \ref{klebladder}.
A ladder with $n$ rungs is denoted as $\mathcal{F}_n$ and can be obtained from $\mathcal{F}_{n-1}$ by acting with the kernel $K$:
\begin{align}
\nonumber
\mathcal{F}_n(t_1,t_2;t_3,t_4) &= \int dt dt'~K(t_1,t_2; t, t') \mathcal{F}_{n-1}(t, t'; t_3,t_4) \, ,\\
\text{where}~~~K(t_1,t_2; t_3,t_4) &= - g_0^2 G(t_1,t_3) G(t_2,t_4) G(t_3,t_4)^2 \, .
\label{kleb_frecursion}
\end{align}
The kernel $K$ commutes with $SL(2,\mathbb{R})$ generators. Given any generator $J$ of $SL(2,\mathbb{R})$, one has
\begin{align}
(J_1 + J_2) K(t_1,t_2;t_3,t_4) &= K(t_1,t_2;t_3,t_4) (J_3 + J_4) \, .
\label{kleb_kcommutes}
\end{align}
Here $J_i$ acts on time $t_i$. Using \refb{kleb_frecursion} one can sum up the ladder diagrams to obtain
\begin{align}
\mathcal{F} &= (1+ K + K^2 + \dots ) \mathcal{F}_0 = \frac{1}{1-K} \mathcal{F}_0 \, . \label{kleb_fsum}
\end{align}
Combining \refb{kleb_kcommutes} with the fact that 
\begin{equation}
\mathcal{F}_0(t_1,t_2;t_3,t_4) \equiv - G(t_1,t_3) G(t_2,t_4) + G(t_1,t_4) G(t_2,t_3)
\end{equation}
preserves the $SL(2,\mathbb{R})$ symmetry, we see that one can use $SL(2,\mathbb{R})$ symmetry to evaluate \refb{kleb_fsum}, although the $K=1$ subspace requires special care and ultimately results in the breaking of the conformal symmetry.

%
\subsection{A class of solvable models} \label{s1.2}
The  KT model can be thought of as a toy model for near extremal black holes. A model which includes  probes of this black hole should have additional fields which preserve the property of ``maximal chaos" and we will present such a model later in \ref{s3}. In this section we investigate some modifications of KT model, which remain solvable at large $N$.

\begin{figure}[H]
	\begin{center}
		\includegraphics[scale=0.5]{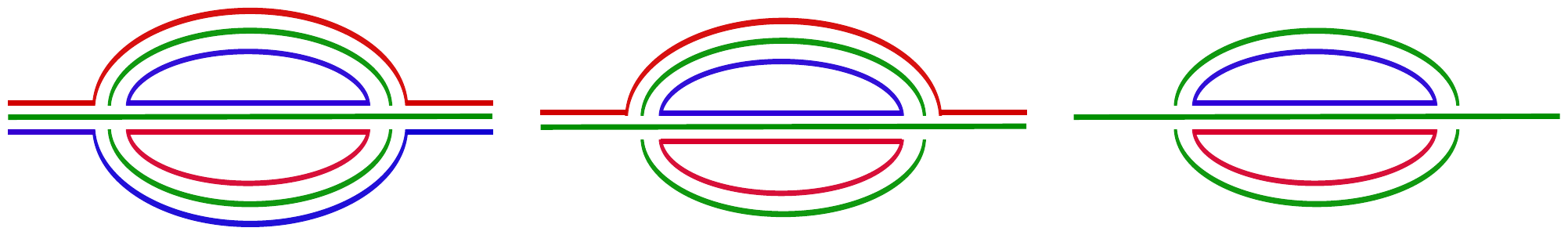}
		\caption{Leading melonic contributions to various propagators.}	\label{uc1_l_melon}
	\end{center}
\end{figure}

We begin with the observation that removing open lines from fig \ref{klebprop} does not change the $N$ dependence. For example removing the open blue line  from fig \ref{klebprop} gives the second diagram of fig. \ref{uc1_l_melon}. This can be thought of as a correction to the propagator for a field carrying two indices (represented by green and red lines) coming from an interaction vertex (see last vertex of fig. \ref{uncolored1_vrtx_ldng}) that can be obtained from fig. \ref{klebfig} by removing one blue line.  This suggests that adding fields carrying less number of indices and interacting through such an interaction leads to new theories with large N structure similar to that of the KT model and therefore also to the SYK model.

\begin{figure}[H]
	\begin{center}
		\includegraphics[scale=0.7]{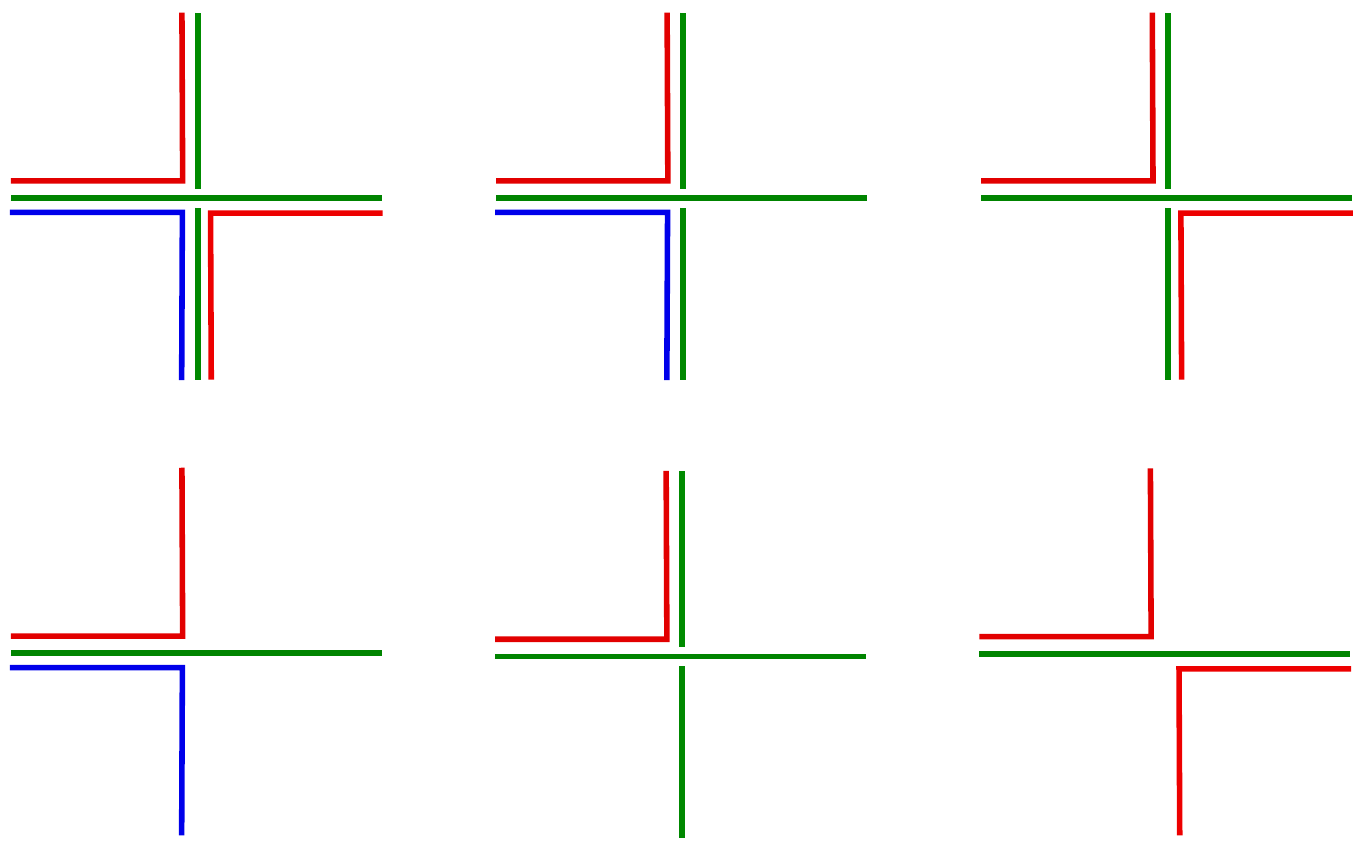}
		\caption{various interaction obtained by removing resolved lines from KT model: red, green and blue resolved lines denote first, second and third resolved indices respectively. E.g. the field $\kappa^{(12)}$ will have one red and one green lines, the field $\eta^{(3)}$ will have one blue line and so on. Permuting colors will give other vertices of same class.}\label{uncolored1}
	\end{center}
\end{figure}

A non exhaustive list of such interaction vertices, obtained from \ref{klebfig} by removing various open lines is given in fig \ref{uncolored1}.
\noindent These interactions include the following new fields:
\begin{equation}
\kappa^{(12)}_{ab}, \kappa^{(23)}_{bc}, \kappa^{(13)}_{ac}, \eta^{(1)}_a, \eta^{(2)}_b, \eta^{(3)}_c \,.
\end{equation}
Here $\kappa^{(ij)}$ carries the $i^{th}$ and $j^{th}$ indices of $\psi$, whereas $\eta^{(i)}$ carries only $i^{th}$ index of $\psi$. This implies that $\kappa^{(ij)}$ transforms as bifundamental under $SO(N)_i \times SO(N)_j$ and trivially under the remaining $SO(N)$ while $\eta^{(i)}$ transforms as fundamental of $SO(N)_i$ and trivially under remaining two $SO(N)$ symmetries.

Among all the interactions of the above kind, only a few are relevant in determining the large $N$ physics. In fig \ref{klebprop}, if one removes any loops, the diagram becomes subleading in $1/N$. 
We claim that only 2 such vertices (up to color permutation) as shown in fig \ref{uncolored1_vrtx_ldng} contribute to the leading order graphs:
\begin{figure}[H]
	\begin{center}
		\includegraphics[scale=0.7]{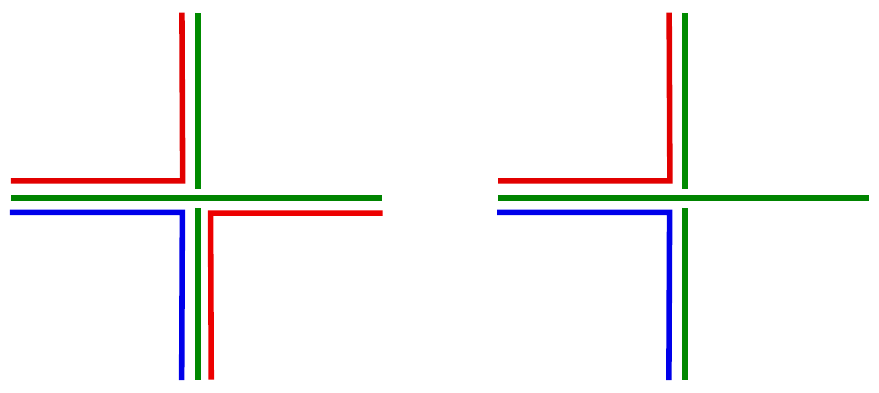}
		\caption{vertices (up to color permutations) relevant for the large $N$ structure}\label{uncolored1_vrtx_ldng}
	\end{center}
\end{figure}
\noindent These interactions, along with original interaction \ref{kleb} give the new leading contributions to various propagators, as shown in fig \ref{uc1_l_melon}. 
\section{Uncolored probe model} \label{s3}
\begin{figure}[H]
	\begin{center}
		\includegraphics[scale=0.5]{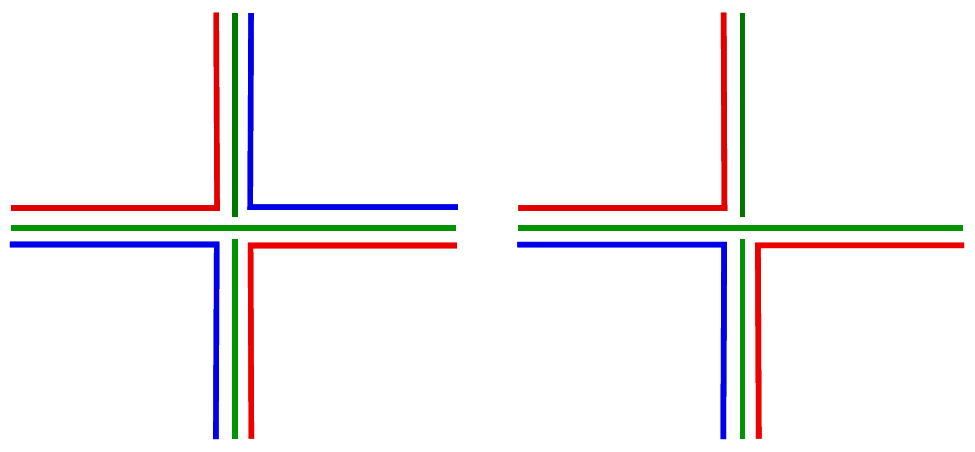}
		\caption{Diagrammatic representation of all the vertices appearing in \ref{our_uc}.}\label{our vertices}
	\end{center}
\end{figure}

In this section we present a class of models motivated by our discussion in section \ref{s1.2} and compute the four point functions and Lyapunov coefficient.  The Hamiltonian for the simplest of our models is given by 
\begin{align}
H  &= \frac{g_0}{4N^{3/2}} \psi_{a_1b_1c_1} \psi_{a_1b_2c_2} \psi_{a_2b_1c_2} \psi_{a_2b_2c_1} + \frac{g_1}{2N^{3/2}}  \psi_{a_1 b_1 c_1} \psi_{a_2b_2c_1} \kappa_{a_1b_2} \kappa_{a_2b_1} 
 \label{our_uc}
\end{align}
where we have introduced factors of $N$ in the interactions such that both $g_0$ and $g_1$ are $\mathcal{O}(N^0)$. Diagrammatic representations of the interaction vertices of \ref{our_uc} are given in fig \ref{our vertices}
We refer to the first term as KT term, which can be thought of as describing dynamics of a black hole, whose effective degrees of freedoms are captured in the field $\psi$. We refer to $\kappa$ (which is same as $\kappa^{(12)}$ in last section) as the ``probe field" and to the second term of \ref{our_uc} as the ``probe term", which is to be thought of as describing the interaction between a black hole and the probe\footnote{KT model has some extra Goldstone modes \cite{Choudhury:2017tax} compared to SYK model which are unaffected by the probe term.}. Interactions involving only probe fields are subleading and not considered in this work.

\begin{figure}[H]
	\begin{center}
		\includegraphics[scale=0.5]{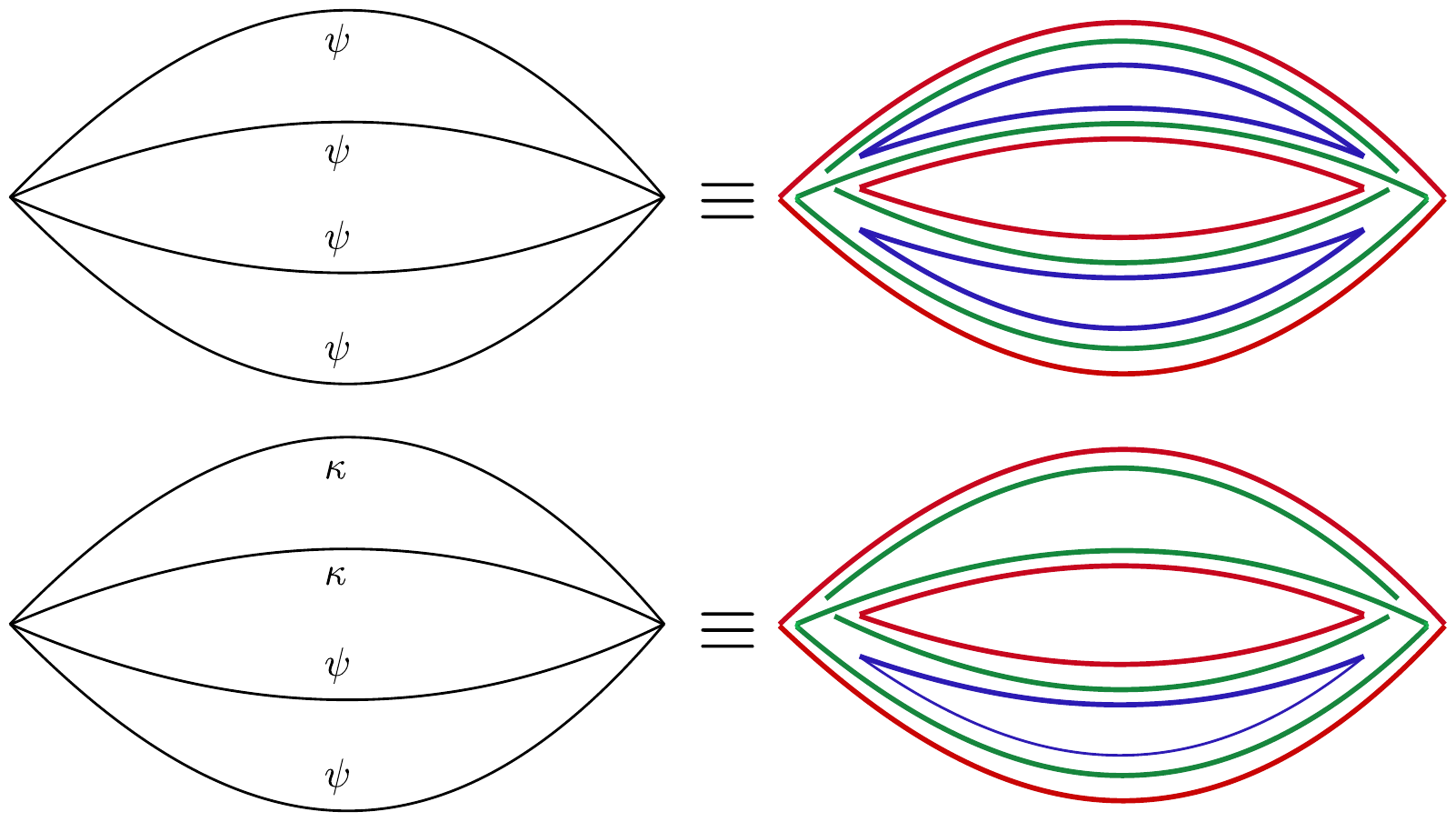}
		\caption{Simplest vacuum diagrams for \refb{our_uc}.}\label{melon_vacuum}
	\end{center}
\end{figure}

\begin{figure}[H]
		\begin{center}
			\includegraphics[scale=0.5]{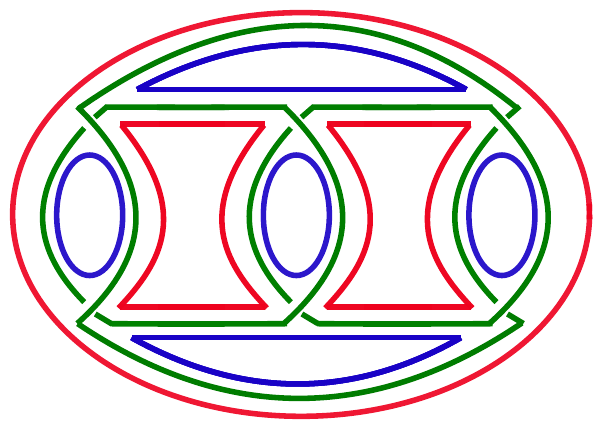} 
			\caption{A subleading bubble in KT theory of order $N^2$}\label{KTN2}
		\end{center}
\end{figure}

The simplest vacuum diagrams for \refb{our_uc} are shown in fig \ref{melon_vacuum}.
\noindent The upper diagram, coming from KT term, scales as $N^3$, whereas the lower diagram, coming from the probe term, scales as $N^2$ and thus gives subleading contributions to free energy. This implies that in the large $N$ limit, thermodynamic properties of the system is entirely determined by the KT term. There are also subleading diagrams arising purely from KT interactions as well. For example the diagram in figure \ref{KTN2} also scales as $N^2$.

\subsection{Propagators}
\paragraph{$G^\psi$:} Propagators for fields $\psi$ is unaffected by the new vertex in the large $N$ limit. This is because the contribution to $G^\psi$ from the probe interaction comes from cutting a $\psi$ line in the lower melon of fig \ref{melon_vacuum}. But this scale as $1/N$ and hence suppressed. Thus in large $N$ limit and in deep infra red, \ref{klebpropa} continues to hold. From now on we will refer to $G^\psi$ simply by $G$. 
\paragraph{$G^\kappa$:}
Simplest contribution to $G^\kappa$ comes from cutting a $\kappa$ line in the lower melon of fig \ref{melon_vacuum}. This removes two loops and the resultant diagram scales as $\cO(N^0)$. Additionally, one can replace an internal $\kappa$ line with such a melon and still get a correction of order $\cO(N^0)$. One can continue doing this and following logic similar to $\psi$ propagators, one gets the following Schwinger Dyson equation for $G^\kappa$
\begin{align}
g_1^2 \int dt G^\kappa (t_1,t) \left( G^\psi(t, t_2)\right)^2 G^\kappa(t,t_2) &= - \delta(t_1 -t_2) \, , \label{SDkappa}
\end{align}
whose solution is
\begin{align}
G^\kappa(t_1,t_2) &= \frac{g_0}{g_1} G^\psi(t_1,t_2) = \frac{bg_0}{g_1} \frac{\sgn(t_1-t_2)}{|t_1 - t_2|^{1/2}} \, . \label{prop_kappa}
\end{align}
We emphasize that $G^\kappa$ is of same order as $G^\psi$ in the $1/N$ expansion but not in a $g_0$ expansion\footnote{Since $b \sim g_0^{-1/2}$ it follows that $G^\kappa \sim g_0^{1/2}$. This dependence is unlike $G$ which scales as $G \sim g_0^{-1/2}$.}. 
At the level of propagators there is an emergent conformal symmetry which is spontaneously broken to $SL(2,\mathbb{R})$.

\subsection{Four point functions}
We now consider four point functions involving various combinations of $\psi$ and $\kappa$ fields. All connected graphs which contribute to these four point functions can be obtained by cutting two lines from appropriate melon diagrams.
\subsubsection{$\langle \psi \psi \psi \psi \rangle$} \label{sec:psi4}
Like the propagator, four point functions of $\psi$ fields are not affected by the presence of new interactions to leading order. Therefore the leading contribution to $\mathcal{F}$, the connected part of $\frac{1}{N^6}\langle \psi_{a_1b_1c_1}(t_1) \psi_{a_1b_1c_1}(t_2) \psi_{a_2b_2c_2}(t_3) \psi_{a_2b_2c_2}(t_4) \rangle$,  comes from summing up the ladder diagrams shown in fig \ref{klebladder} and is given by \refb{kleb_fsum}, which we repeat for convenience\footnote{From now on, we will use $K$ for $K^{\psi}$, $\mathcal{F}$ for $\mathcal{F}^{\psi}$ and $\mathcal{F}_0$ for $\mathcal{F}_0^{\psi}$.}
\bea
\mathcal{F} &=& (1+ K + K^2 + \dots ) \mathcal{F}_0 = \frac{1}{1-K} \mathcal{F}_0 , \label{F1overKF0}\\
\text{where,}~~ K(t_1,t_2; t_3, t_4) &=& -g_0^2 G (t_1,t_3) G (t_2,t_4) G(t_3,t_4)^2,\non \\
\mathcal{F}_0(t_1,t_2;t_3,t_4) &=& - G(t_1,t_3) G(t_2,t_4) + G(t_1,t_4) G(t_2,t_3) \, . \label{fpsi}
\eea
Following  \cite{Maldacena:2016hyu} we define in the conformal limit, the normalized four point function:
\begin{align}
\frac{\mathcal{F}(t_1,t_2;t_3,t_4) }{G(t_{12}) G(t_{34}) }&= \mathcal{F}(\chi) \,,\qquad \text{where}\qquad\chi= \frac{t_{12} t_{34}}{t_{13} t_{24}} \, .\label{Fchidefin}
\end{align}
and note that $\mathcal{F}(\chi)$ was evaluated in \cite{Maldacena:2016hyu}. 

A central point of \cite{Maldacena:2016hyu} is that in the strict conformal limit, the four point function of the SYK-model diverges. In addition to the finite and conformally invariant component of the four point function $\cF(\chi)$, there is an additional non-trivial component $\mathcal{F}(t_1,t_2;t_3,t_4)$ which diverges and breaks conformal symmetry (thus cannot be expressed as a function of $\chi$ alone). This non-conformal component of $\cF^\psi$ is denoted $\cF_{h=2}^\psi$ due to  the manner in which the eigenvalues of $K$ are parameterized in \cite{Maldacena:2016hyu}. The unit eigen-subspace of $K$ corresponds to $h=2$ and from \eq{F1overKF0} we see that it is this subspace which leads to the divergence. One proceeds by regulating the spectrum of $K$ and the first non-trivial correction to the unit eigenvalue of $K$ is of order $\frac{1}{\beta g_0}$, which in turn gives a contribution of order 
\be
{\cal F}_{h=2}^\psi \sim \beta g_0 +\ldots  \label{Fpsih2}
\ee
where the ellipsis represent lower order terms in the expansion in $\beta g_0 $. Subject to certain assumptions, the first subleading term in \eq{Fpsih2} has been computed in \cite{Maldacena:2016hyu}.

For our later discussion of the spectrum, it will be useful to mention that in the short time limit $\chi \rightarrow 0$ we have
\begin{align}
{\cal F}(\chi) &= \sum_{m=1}^\infty c_m^2 \chi^{h_m} \, ,~\text{where,}~~c_m^2= \alpha_0 \frac{(h_m - 1/2)}{\pi \tan (\pi h_m /2) k'(h_m)} \frac{\Gamma(h_m)^2}{\Gamma(2h_m)} \, , \label{fchi}
\end{align}
where $\alpha_0 = g_0^2 b^4, b= (4 \pi g_0^2)^{-1/4}$ and $h_m$ is the $m^{th}$ root\footnote{We define $h_0=2$} of the equation 
\be
3\tan \frac{\pi (h-1/2)}{2}= 1-2h\,.\ee

\subsubsection{$\langle \kappa \kappa \kappa \kappa \rangle$} \label{skappa4}
\begin{figure}[H]
	\begin{center}
		\includegraphics[scale=0.5]{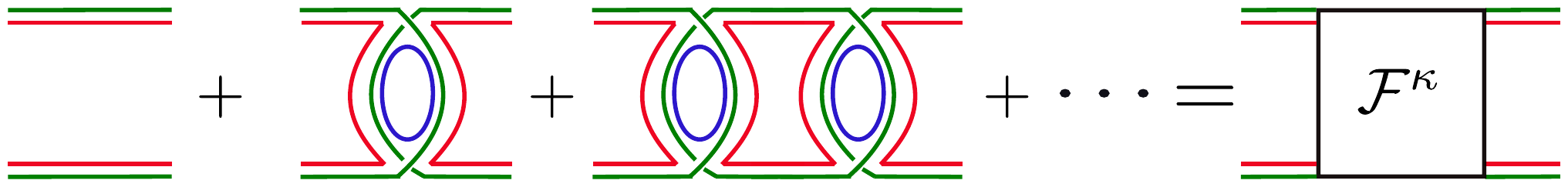}
		\caption{Diagrams contributing to 4 point function of $\kappa$}\label{kappa4}
	\end{center}
\end{figure}

The gauge invariant four point function of the probe fields has the following structure:
\begin{align}
\frac{1}{N^4} \sum_{a_1,b_1,a_2,b_2}\langle \kappa_{a_1b_1}(t_1) \kappa_{a_1b_1}(t_2) \kappa_{a_2b_2}(t_3) \kappa_{a_2b_2}(t_4) \rangle &= G^\kappa(t_1,t_2) G^\kappa(t_3,t_4) + \frac{1}{N^2} \mathcal{F}^{\kappa}(t_1,t_2;t_3,t_4)\label{kappa4defn}
\end{align}
and $\mathcal{F}^\kappa$ is given by sum of ladders shown in fig \ref{kappa4}. The first connected piece of $\mathcal{F}^\kappa$ is obtained by cutting two $\kappa$ lines in the lower melon of fig \ref{melon_vacuum}. Subsequent diagrams can be obtained by cutting more complicated melons although it is easier to think of these ladder diagrams as being obtained by successively adding rungs constructed only from $\psi$-fields.
The sum of ladders in fig \ref{kappa4} has the same structure as $\mathcal{F}^\psi$ and one gets
\begin{align}
\mathcal{F}^\kappa &= \frac{1}{1- K^\kappa} \mathcal{F}^\kappa_0 = \frac{g_0^2}{g_1^2} \mathcal{F}^\psi \, . \label{fkappa}
\end{align}
where we have used
\begin{align*}
K^\kappa(t_1,t_2; t_3, t_4) &= -g_1^2 G^\kappa (t_1,t_3) G^\kappa (t_2,t_4) G^\psi(t_3,t_4)^2 =  K^\psi(t_1,t_2; t_3, t_4) ,\\
\mathcal{F}^\kappa_0(t_1,t_2;t_3,t_4) &= - G^\kappa(t_1,t_3) G^\kappa(t_2,t_4) + G^\kappa(t_1,t_4) G^\kappa(t_2,t_3) = \frac{g_0^2}{g_1^2} \mathcal{F}^\psi_0(t_1,t_2;t_3,t_4) \, .
\end{align*}
Similarly to \eq{Fchidefin} we define in the strict conformal limit
\begin{align}
\mathcal{F}^\kappa(t_1,t_2;t_3,t_4) &= G^\kappa(t_{12}) G^\kappa(t_{34}) \mathcal{F}^\kappa(\chi) \, 
\end{align}
and then \refb{fkappa} along with \refb{prop_kappa} implies that the conformally invariant component of the four point function has the same normalization as ${\cal F}^{\psi}$:
\begin{align}
\mathcal{F}^\kappa(\chi) &= \mathcal{F}(\chi) \, . \label{fkappa=f}
\end{align}
From \refb{fkappa} we see that $\mathcal{F}^\kappa$ has a conformally invariant part $\mathcal{F}^\kappa_{h \neq 2}$ and a component $\mathcal{F}^\kappa_{h=2}$ which spontaneously breaks conformal symmetry. In addition, as for $\cF^\psi$ the  Lyapunov coefficient is maximal. Interestingly, \refb{fkappa} implies that the probe fields make a copy of the spectrum of primaries of the  theory with just the $\psi$-fields. 

\subsubsection{$\langle \psi \psi \kappa \kappa \rangle$} \label{sec:psipsikapkap}
\begin{figure}[h]
	\begin{center}
		\includegraphics[scale=0.5]{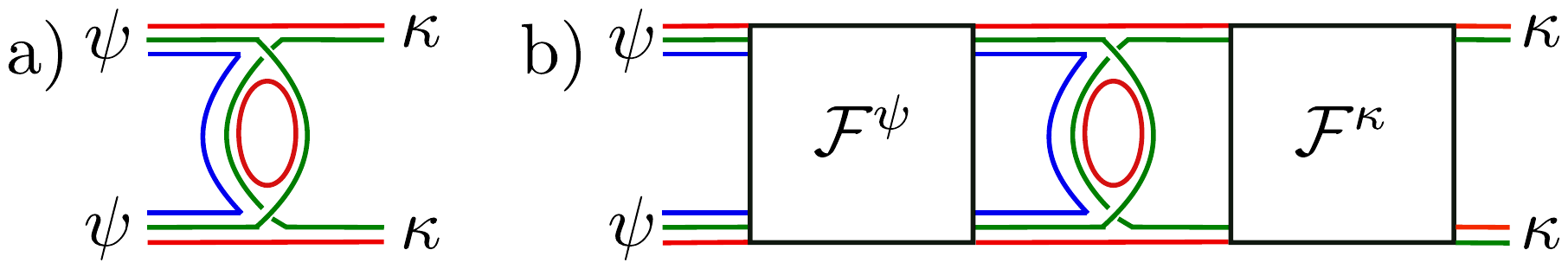}
		\caption{a) Basic vertex contributing to $\langle \psi \psi \kappa \kappa\rangle$. \quad b) One can continue adding rungs on both side of the first diagram to obtain this structure at all orders}\label{psi2kappa2}
	\end{center}
\end{figure}

The mixed four point function is
\begin{align}
\frac{1}{N^5} \sum_{a_1,b_1,c_1,a_2,b_2} \langle \psi_{a_1b_1c_1}(t_1) \psi_{a_1b_1c_1}(t_2) \kappa_{a_2b_2}(t_3) \kappa_{a_2b_2}(t_4) \rangle &= G^\psi(t_1,t_2) G^\kappa(t_3,t_4) + \frac{1}{N^3} \mathcal{F}^{\psi \kappa} \, .\label{psi2kappa2defn}
\end{align}
 The simplest contribution to $\mathcal{F}^{\psi \kappa}$ comes from a diagram obtained by cutting a $\psi$ line and a $\kappa$ line in the lower melon of fig \ref{melon_vacuum}. This is a ladder with a single rung as shown in fig \ref{psi2kappa2}\textcolor{blue}{a}, where the uncontracted lines on the left correspond to $\psi$ fields and thus have three unresolved components while those on the right correspond to $\kappa$ fields and have two unresolved lines. We can express it as
\begin{align}
\frac{g_1}{g_0} \int dt dt'~K(t_1,t_2; t,t') \mathcal{F}_0^\kappa(t,t';t_3,t_4) &= \frac{g_1}{g_0} \left( K * \mathcal{F}_0^\kappa \right)(t_1,t_2;t_3,t_4) \, .
\end{align}

Now to generate the set of graphs which contribute to $\cF^{\psi\kappa}$ at leading order in the $1/N$ expansion, one should add appropriate rungs on both sides of the given rung as presented in figure \ref{psi2kappa2}\textcolor{blue}{b}. All the rungs on both the left and the right correspond to $\psi$-fields being exchanged as do all the side-rails except for two internal rails and the two uncontracted lines on the right, which represent $\kappa$-fields.
The final result follows immediately from figure \ref{psi2kappa2}\textcolor{blue}{b} is 
\begin{align}
\mathcal{F}^{\psi \kappa} &= \frac{g_0}{g_1} \frac{K}{(1-K)^2} \mathcal{F}_0 \, . \label{FKF0kappa}
\end{align}

Similarly as for $\cF^\psi$  we have a component which depends only on the conformal cross ration $\chi$ and an additional component which breaks conformal symmetry:
\begin{align}
\frac{\mathcal{F}^{\psi \kappa}(t_1, t_2, t_3, t_4)}{G^\kappa(t_{12}) G(t_{34}) } &= \mathcal{F}_{h\neq2}^{\psi\kappa}(\chi) + \mathcal{F}^{\psi \kappa}_{h= 2}(t_1, t_2, t_3, t_4)\, .
\end{align}
The conformal part  $\mathcal{F}_{h\neq2}^{\psi \kappa}(\chi)$ has a very similar structure to the conformal component of $\mathcal{F}^\psi$ or $\mathcal{F}^\kappa$ and we evaluate this in appendix \ref{fourdetails}. Since the kernel $K$ commutes with the generators of the conformal group, this computation utilizes essentially the same techniques as used in the computation of the conformal part of $\cF^{\psi}$.

To regulate the divergence in the non-conformal $h=2$ subspace one must compute the four point function taking into account broken conformal symmetry and further corrections in $\frac{1}{\beta g_0}$ and $\frac{1}{\beta g_1}$. It remains a difficult task to precisely evaluate  $\mathcal{F}^\kappa_{h=2}$ along the lines of the strategy in \cite{Maldacena:2016hyu} for computing $\mathcal{F}^\psi_{h=2}$ and in the current work will only note the leading scaling behavior. Following a similar argument to that which we outlined in section \ref{sec:psi4} for the leading correction to $\cF^\psi_{h=2}$, the double pole in \eq{FKF0kappa} implies that after regulating the eigenspace of $K$, the leading contribution to $\mathcal{F}^\kappa_{h=2}$ scales as
\be
\mathcal{F}^{\psi \kappa}_{h=2}(t_1, t_2, t_3, t_4) \sim (\beta g_0)^2 +\ldots\, ,
\ee
which is a higher scaling in $\beta g_0$ that the leading contribution to $\cF^\psi_{h=2}$ of $\cF^{\kappa}_{h=2}$.

\subsection{Chaos} \label{chaosection}
In a chaotic quantum system, out of time order correlation functions grow exponentially\footnote{For the SYK model, this growth is followed by an exponential decay \cite{Maldacena:2016upp} and then a power law decay \cite{Bagrets:2017pwq}.}. In the present case, there are three out of time order correlators that one may look at in order to diagnose chaos, namely
\bea
F^\psi (t_1,t_2) &= \Tr{} \left[ y\, \psi_{abc}(t_1) \psi_{a'b'c'}(0)y \, \psi_{abc}(t_2) \psi_{a'b'c'}(0)   \right] \non \\
F^\kappa (t_1,t_2) &= \Tr{} \left[ y\, \kappa_{ab}(t_1) \kappa_{a'b'}(0)y \, \kappa_{ab}(t_2) \kappa_{a'b'}(0)   \right] \\
F^{\psi \kappa} (t_1,t_2) &= \Tr{} \left[ y \, \psi_{abc}(t_1) \kappa_{a'b'c'}(0)y \, \psi_{abc}(t_2) \kappa_{a'b'}(0)   \right] \, , \non
\eea
where, $y= \rho(\beta)^{1/4}$, $\rho(\beta)$ being the density matrix at inverse temperature $\beta$ and repeated indices are summed over.

The out of time ordered correlators $F^\psi$ and $F^\kappa$ have the same ladder structure as corresponding correlator in SYK model. In both correlators, rungs are given by
\begin{align}
K_R(t_1,t_2;t_3,t_4) &= g_0^2 G_R(t_1,t_3) G_R(t_2,t_4) G_{lr}(t_3,t_4)^2 \, ,
\end{align}
where $G_R$ is the retarded Green's function and $G_{lr}$ is a Wightman correlator:
\begin{align}
G_R(t) &= 2b \cos (\pi \Delta) \theta(t) \left[ \frac{\pi}{\beta \sinh \frac{\pi t}{\beta}} \right]^{2\Delta}~, ~G_{lr}(t) = b\left[ \frac{\pi}{\beta \sinh \frac{\pi t}{\beta}} \right]^{2\Delta} \, .
\end{align}
When $F^\psi$ (or $F^\kappa$) is acted upon by $K_R$, one gets back $F^\psi$ (or $F^\kappa$), except the $0^{th}$ piece. But this piece has a $G_R(t_1) G_R(t_2)$ in it, which is negligible for large $t_1,t_2$. So, in this limit, $F^\psi$ (or $F^\kappa$) is an eigenfunction of $K_R$ with eigenvalue $1$. In order to solve for this, one can make an ansatz (we will write for $F^\psi$, $F^\kappa$ has same expression up to an over all factor of $g_0^2/g_1^2$.)
\begin{align}
F^\psi(t_1,t_2) &= e^{\lambda_L^\psi (t_1+t_2)/2}f(t_{12}) \, , \label{lyapunovdefn}
\end{align} 
$\lambda_L^\psi$ being the Lyapunov coefficient for $F^\psi$ (which is same as $\lambda_L^\kappa$, the Lyapunov coefficient for $F^\kappa$). Looking at eigenfunctions of $K_R$, one finds they are given by
\begin{align}
\ell_h(t_1,t_2) &= \frac{e^{-h\frac{\pi}{\beta}(t_1 + t_2)}}{\left[ \cosh \frac{\pi}{\beta} t_{12} \right]^{1/2-h}}
\label{kreigen}
\end{align} 
with eigenvalue
\begin{align}
k_R(h) &= \frac{\Gamma(5/2) \Gamma(1/2-h)}{\Gamma(3/2) \Gamma(3/2-h)} \, ,
\end{align}
in the conformal limit. It is then easy to see that only solution for $k_R(h)=1$ is $h=-1$. In large $t_1,t_2$, one has
\begin{align}
F^\psi(t_1,t_2) &= \frac{e^{\pi (t_1+t_2)/\beta}}{\left[ \cosh \frac{\pi}{\beta} t_{12} \right]^{1/2-h}} \, .
\end{align}
Comparing with \refb{lyapunovdefn}, we see
\begin{align}
\lambda_L^\psi &= 2\pi/\beta \, ,
\end{align}
which is maximal Lyapunov coefficient\cite{Maldacena:2015waa} and black holes are known to saturate this bound. One also has $\lambda^\kappa_L= 2\pi/\beta$ along similar lines, thus mimicking the chaotic dynamics of Hawking radiation.

For the remaining correlator $F^{\psi \kappa}$, a slightly modified version of the above argument holds. We define a new kernel
\be
\hK_R=(K_R+K_R^{-1}-1)\,,
\ee
then observe that  in the limit $t_1,t_2 >>1$
\be
\hK_R \, F^{\psi \kappa}(t_1,t_2)=F^{\psi \kappa}(t_1,t_2)\,. \label{hKRFpsikappa}
\ee
Thus in this limit $F^{\psi \kappa}(t_1,t_2)$ is an eigenfunction of $\hK_R$, with eigenvalue $1$. Now since $\hK_R$ commutes with $K_R$, we deduce from \eq{hKRFpsikappa} that 
\be
K_R \, F^{\psi \kappa}(t_1,t_2)=F^{\psi \kappa}(t_1,t_2)\,. \label{KRFpsikappa}
\ee
Thus the desired eigenfunction is essentially $\ell_{-1}(t_1,t_2)$ given in \refb{kreigen} and $F^{\psi \kappa}$ also has maximal Lyapunov coefficient $\lambda_L^{\psi \kappa}=2\pi/ \beta$.

\subsection{Spectrum}
The SYK model (and its tensorial cousins) are widely believed to constitute an example of so called $NAdS_2/NCFT_1$  \cite{Maldacena:2016hyu}. Conformal primaries of the SYK model correspond to bulk fields whose masses are determined by the scaling dimensions of the primaries. To compute these, one first expresses the gauge invariant fermion bilinear as (we write it for the KT model)
\begin{align}
\sum_{a,b,c} \psi_{abc}(t_1) \psi_{abc}(t_2) &= N^{3/2} G(t_{12}) \sum_n c_{\psi, n} |t_{12}|^{h'_n} \mathcal{O}_{\psi, n} \left( \frac{t_1 + t_2}{2} \right) \, ,
\end{align}
where $h'_n$ is the conformal dimension of $n^{th}$ primary, i.e. 
\begin{align}
\langle \mathcal{O}_{\psi, m} (t_1) \mathcal{O}_{\psi, n} (t_2) \rangle &= \frac{\delta_{mn}}{|t_{12}|^{2h'_n}} \, .
\end{align}.
To compute $h'_n$-s and $c_{\psi,n}$-s, one notes in the limit $t_{12} \rightarrow 0, t_{34} \rightarrow 0$, 
\begin{align}
\nonumber
&\sum_{a,b,c; \atop a',b',c'} \langle \psi_{abc}(t_1) \psi_{abc}(t_2) \psi_{a'b'c'}(t_3) \psi_{a'b'c'}(t_4) \rangle \\
\nonumber
&= N^3 G(t_{12}) G(t_{34}) \sum_{m,n} c_{\psi, m} c_{\psi, n} |t_{12}|^{h'_m} |t_{34}|^{h'_n} \langle \mathcal{O}_{\psi, m} \left( \frac{t_1 + t_2}{2} \right) \mathcal{O}_{\psi, n} \left( \frac{t_3 + t_4}{2} \right) \rangle \\
&=  N^3 G(t_{12}) G(t_{34}) \sum_{n} c_{\psi, n}^2 \frac{ |t_{12}|^{h'_n} |t_{34}|^{h'_n}}{|t_{13}|^{2h'_n}} \, .
\label{o21}
\end{align}
Again in the same limit, one has
\begin{align}
\sum_{a,b,c; \atop a',b',c'} \langle \psi_{abc}(t_1) \psi_{abc}(t_2) \psi_{a'b'c'}(t_3) \psi_{a'b'c'}(t_4) \rangle  &= N^3G(t_{12}) G(t_{34}) \sum_{m=1}^\infty c_m^2 \left( \frac{t_{12} t_{34}}{t_{13} t_{24}} \right)^{h_m} \, ,
\label{o22}
\end{align}
where   $c_m^2$ are given in \refb{fchi}. Comparing \refb{o21} and \refb{o22}, one has 
\begin{align}
h'_{\psi, n} &= h_n,~c_{\psi,n} = c_n \, . 
\end{align}
In the dual bulk theory, corresponding to a boundary primary of dimension $h_n$ one gets a bulk field $\phi_n$ of mass $m_n^2=h_n(h_n-1)$.

In the present case, there are extra primaries due to $\kappa$-fields. We restrict to the simplest case of \refb{our_uc}. Then similarly one gets
\begin{align}
\sum_{a,b} \kappa_{ab}(t_1) \kappa_{ab}(t_2) &= N G^\kappa(t_{12}) \sum_{n} c_n |t_{12}|^{h_n} \mathcal{O}_{\kappa,n} \left( \frac{t_1 + t_2}{2}\right) \, .
\end{align}
These primaries lead to a new set of bulk fields $\phi'_n$ with same masses $m_N^2=h_n(h_n-1)$. 

To leading order we have $\langle \mathcal{O}_{m,\psi} \mathcal{O}_{n,\kappa} \rangle =0~\forall m,n$. Thus for every mass level $n$, there is an independent $U(1)$ symmetry that rotates $\mathcal{O}_{n,\psi}$ and $\mathcal{O}_{n,\kappa}$. This symmetry is broken though by the following $\mathcal{O}(1/\sqrt{N})$ correction\footnote{At this order there are no other correction, since NLO corrections \cite{Dartois:2017xoe} to $\mathcal{F}$ are off by a factor of $1/N$.} coming from the $\langle \psi^2 \kappa^2 \rangle$ four point function:
\begin{align}
 \langle \mathcal{O}_{\psi,m}(t_1) \mathcal{O}_{\kappa,n}(t_2) \rangle &= \frac{\widetilde{c}_m^2}{c_m^2 \sqrt{N}} \frac{\delta_{mn}}{|t_{13}|^{2h_n}} \, .
\end{align}
Now one can choose the following combination of primaries, that are orthonormal up to this order:
\begin{align}
\nonumber
\mathcal{O}_{m+} &= \left[ 2+ \frac{2 \tilde{c}_m^2}{\sqrt{N} c_m^2} \right]^{-1/2} \left( \mathcal{O}_{\psi,m} + \mathcal{O}_{\kappa,m} \right) \\
\mathcal{O}_{m-} &= \left[ 2 - \frac{2 \tilde{c}_m^2}{\sqrt{N} c_m^2} \right]^{-1/2} \left( \mathcal{O}_{\psi,m} - \mathcal{O}_{\kappa,m} \right) \, .
\end{align}

\subsection{Adding additional fields} \label{adding more fields}
In section \ref{s1.2}, we considered a general class of models which are are obtained by adding additional fields in KT model but remain solvable at large $N$ in the deep infrared. We then considered a particular case in section \ref{s2}. Here we shed some light on the general case.

First we restore permutation symmetry between the three copies of $SO(N)$. This amounts to introducing three new fields to the original KT model: $\kappa^{(12)}, \kappa^{(13)}, \kappa^{(23)}$. This is implemented by replacing the probe term in \refb{our_uc} by the following term
\begin{align}
V_1 &= \frac{g_1}{2N^{3/2}} \left( \psi_{a_1 b_1 c_1} \psi_{a_2b_2c_1} \kappa^{(12)}_{a_1b_2} \kappa^{(12)}_{a_2b_1} +  \psi_{a_1 b_1 c_1} \psi_{a_2b_1c_2} \kappa^{(13)}_{a_2 c_1} \kappa^{(13)}_{a_1c_2} + \psi_{a_1b_1c_1} \psi_{a_1 b_2 c_2} \kappa^{(23)}_{b_1c_2} \kappa^{(23)}_{b_2 c_1} \right) \, .
\end{align}
It is easy to check that $G^{\kappa^{12}}=G^{\kappa^{13}}=G^{\kappa^{23}}=G^\kappa$, where $G^\kappa$ is same as \refb{prop_kappa}.

\begin{figure}[H]
	\begin{center}
		\includegraphics[scale=0.5]{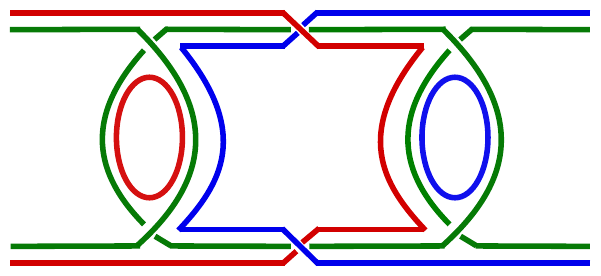}
		\caption{Simplest leading contribution to $ \langle \kappa^{(12)} \kappa^{(12)} \kappa^{(23)} \kappa^{(23)} \rangle$}\label{kappa4_hetero}
	\end{center}
\end{figure}

Coming to four point functions, again it is easy to check that $\langle \psi \psi \kappa^{(ij)} \kappa^{(ij)} \rangle$ is the same for all three $(ij)$-s and has the same expression as in \refb{psi2kappa2defn}. Similarly the four point function $\langle \kappa^{(ij)} \kappa^{(ij)} \kappa^{(ij)} \kappa^{(ij)} \rangle$ is also same for all three $(ij)$-s and same as \refb{kappa4defn}. Along with these, now there is a new four point function $\langle \kappa^{(ij)} \kappa^{(ij)} \kappa^{(jk)} \kappa^{(jk)} \rangle$. This is the same for all three choices of $(ij),(jk)$ by permutation symmetry. As an example, we consider the case $(ij)=(12), (jk)=(23)$. It has the following structure
\begin{align}
\frac{1}{N^4} \sum_{a,b, b',c'} \langle \kappa^{(12)}_{ab}(t_1) \kappa^{(12)}_{ab}(t_2) \kappa^{(23)}_{b'c'}(t_3) \kappa^{(23)}_{b'c'}(t_4) \rangle &= G^\kappa(t_{12}) G^\kappa(t_{34}) + \frac{1}{N^3} \widetilde{\mathcal{F}}^\kappa \, .
\end{align}
and the simplest contribution to $\widetilde{\mathcal{F}}^\kappa$ is shown in figure \ref{kappa4_hetero}.
One can now continue adding rungs in the left, right and center to generate other contributions to $\widetilde{\mathcal{F}}^\kappa$. The final answer is
\begin{align*}
\widetilde{\mathcal{F}}^\kappa &= \frac{g_0^2}{g_1^2} \frac{K^2}{(1-K)^3} \mathcal{F}_0
\end{align*}
and the conformal part of $\widetilde{\mathcal{F}}^\kappa$ is computed in \ref{fourdetails}. Conformal symmetry is broken by $h=2$ subspace. Following the argument of section \ref{sec:psipsikapkap}, we note that
\begin{align}
\frac{\widetilde{\mathcal{F}}^\kappa_{h=2}(t_1,t_2,t_3,t_4)}{G^\kappa(t_{12}) G^\kappa(t_{34})} &\sim (\beta g_0)^3 \, .
\end{align}
Generalizing the chaos computations of section \ref{chaosection}, it follows that the corresponding OTOC
\begin{align}
\widetilde{F}^\kappa(t_1,t_2) &= \Tr{} \left[ y \kappa^{(ij)} (t_1) y \kappa^{(ij)}(0) y \kappa^{(ij)} (t_2) y \kappa^{(ij)}(0) \right] \, ,
\end{align}
also has maximal Lyapunov coefficient $\lambda_L= 2\pi /\beta$.

Next, one can add fields carrying a single index. This corresponds to introducing the following interaction in the Hamiltonian:
\begin{align}
V_2 &= \frac{g_2}{N^{3/2}} \left( \psi_{a_1 b_1 c_1} \kappa^{(23)}_{b_2 c_1}  \kappa^{(12)}_{a_1b_2} \eta^{(2)}_{b_1}  + \psi_{a_1 b_1 c_1} \kappa^{(13)}_{a_1c_2} \kappa^{(23)}_{b_1 c_2} \eta^{(3)}_{c_1} + \psi_{a_1 b_1 c_1} \kappa^{(12)}_{a_2 b_1} \kappa^{(13)}_{a_2 c_1} \eta^{(1)}_{a_1} \right)  \, .
\end{align}
Starting with propagators, first we note all three $G^{\eta_i}~(i=1,2,3)$ are the same by permutation symmetry and we will call them $G^\eta$. The Schwinger Dyson equation for $G^\eta $ turns out to be very similar to $G^\psi$ and $G^\kappa$ and has the following solution
\begin{align}
G^\eta(t) &= \frac{g_1^2}{g_2^2} G(t) = \frac{bg_2^2}{g_1^2} \frac{\sgn(t)}{|t|^{1/2}} \, .
\end{align} 
The four point functions involving only $\psi$ and $\kappa$ are not affected by $V_2$ to leading order. There are quite a few new four point functions involving $\eta$. It can be checked that they all show the same pattern of conformal symmetry breaking and exhibit maximal chaos.

\section{Colored probe model} \label{s4}

We first briefly recall the Gurau-Witten model \cite{Witten:2016iux}. For $D+1$ colors, the model contains $D+1$ Majorana fermions $\psi^0,  \psi^1, \dots, \psi^D$ and one associates a global $SO(N)$ symmetry to each pair $(ij)$: a fermion $\psi^i$ sits in fundamental of $SO(N)_{ij}~\forall j \neq i$. Thus each fermion carries $D$ indices and each index runs from $1$ to $N$. For example, writing all the indices of $\psi^0$, one would have $\psi^0_{i_1 i_2 \dots i_D}$ and so on.  The Hamiltonian for this model is given by 
\begin{align}
H_{GW} &= \frac{g_0}{N^{D(D+1)/4}} \psi^0 \psi^1 \dots \psi^D \, , \label{Hgw}
\end{align} 
where we have suppressed the indices. At leading order in $1/N$ this model leads to same physics as SYK model. Further it had the advantage over SYK model of being fully quantum mechanical. One can further ``uncolor" \cite{Bonzom:2012hw} this model to get the KT model \cite{Klebanov:2016xxf}.

Now we add probe fields to this model. We would denote a probe field, obtained from say $\psi^0$ by removing the index $i_{01}$, as $\psi^{0 \widehat{1}}$. Similarly $\psi^{2 \widehat{0} \widehat{1}}$ denotes a field obtained from $\psi^{2}$ by removing indices $i_{02}, i_{12}$.  Since we have already gathered some understanding of probe fields in \refb{s3}, we add all possible probe fields at one go. This amounts to adding the following interactions with \ref{Hgw}:
\begin{align}
\nonumber
V_1 &= \frac{g_1}{2! (D-1)! N^{D(D+1)/4}} \epsilon_{c_0 \dots c_D} \psi^{c_0 \widehat{c_1}} \psi^{c_1 \widehat{c_0}} \psi^{c_2} \dots \psi^{c_D} \, ,\\
\nonumber
V_2 &= \frac{g_2}{2! (D-2)! n^{D(D+1)/4}} \epsilon_{c_0 \dots c_D} \psi^{c_0 \widehat{c_1} \widehat{c_2}} \psi^{c_1 \widehat{c_0}} \psi^{c_2 \widehat{c_0}} \psi^{c_3} \dots \psi^{c_D}  \, ,\\
\nonumber
&\vdots \\
\nonumber
V_p &= \frac{g_p}{p! (D-p)! n^{D(D+1)/4}} \epsilon_{c_0 \dots c_D} \psi^{c_0 \widehat{c_1} \dots \widehat{c_p}} \psi^{c_1 \widehat{c_0}} \dots \psi^{c_p \widehat{c_0}} \psi^{c_{p+1}} \dots \psi^{c_D} \, ,\\
\nonumber
&\vdots \\
V_{D-1} &= \frac{g_{D-1}}{(D-1)! n^{D(D+1)/4}} \epsilon_{c_0 \dots c_D} \psi^{c_0 \widehat{c_1} \dots \widehat{c}_{D-1}} \psi^{c_1 \widehat{c_0}} \dots \psi^{c_{D-1} \widehat{c_0}} \psi^{c_D}\, , \label{colorprobe}
\end{align}
where $c_1 , \dots, c_D$ run from $0$ to $D$. The numerical prefactors are chosen to ensure that every term appears only once. E.g. for $D=3$, one has 
\begin{align}
\nonumber
V_1 &= \frac{g_1}{N^{3/2}} \left[ \psi^{0, \widehat{1}} \psi^{1, \widehat{0}} \psi^{2} \psi^{3} + \psi^{0, \widehat{2}} \psi^{1} \psi^{2, \widehat{0}} \psi^{3} + \psi^{0, \widehat{3}} \psi^{1} \psi^{2} \psi^{3 \widehat{0}} + \psi^0 \psi^{1 \widehat{2}} \psi^{2 \widehat{1}} \psi^3 + \psi^0 \psi^{1 \widehat{3}} \psi^2 \psi^{3 \widehat{1}} + \psi^0 \psi^1 \psi^{2 \widehat{3}} \psi^{3 \widehat{2}} \right] \\
\nonumber
V_2 &= \frac{g_2}{N^{3/2}} \big[ 
\psi^{0 \widehat{1} \widehat{2}} \psi^{1 \hat{0}} \psi^{2 \widehat{0}} \psi^{3} 
+ \psi^{0 \widehat{2} \widehat{3}} \psi^{1} \psi^{2 \widehat{0}} \psi^{3 \widehat{0}} 
+ \psi^{0 \widehat{1} \widehat{3}} \psi^{1 \widehat{0}} \psi^{2} \psi^{3 \widehat{0}} 
+ \psi^{0 \widehat{1}} \psi^{1 \widehat{0} \widehat{3}} \psi^2 \psi^{3 \widehat{1}} \\
\nonumber
&+ \psi^{0 \widehat{1}} \psi^{1 \widehat{2} \widehat{0}} \psi^{2 \widehat{1}} \psi^3
+ \psi^0 \psi^{1 \widehat{3} \widehat{2}} \psi^{2 \widehat{1}} \psi^{3 \widehat{1}}
+ \psi^{0 \widehat{2}} \psi^{1 \widehat{2}} \psi^{2 \widehat{0} \widehat{1}} \psi^3
+ \psi^{0 \widehat{2}} \psi^1 \psi^{2 \widehat{3} \widehat{0}} \psi^{3 \widehat{2}} \\
&+ \psi^0 \psi^{1 \widehat{2}} \psi^{2 \widehat{1} \widehat{3}} \psi^{3 \widehat{2}} 
+ \psi^{0 \widehat{3}} \psi^1  \psi^{2 \widehat{3}} \psi^{3 \widehat{2} \widehat{0}} + \psi^{0 \widehat{3}} \psi^{1 \widehat{3}} \psi^2 \psi^{3 \widehat{1} \widehat{0}}  + \psi^0 \psi^{1 \widehat{3}} \psi^{2 \widehat{3}} \psi^{3 \widehat{1} \widehat{2}} 
\big] \, ,
\end{align}
and so on. The analysis of four point functions is very similar to those of uncolored probe model and exhibit same pattern of conformal symmetry breaking as well as maximal chaos.

\section{Discussion and Future Directions} \label{s5}
In this paper we have presented various tensor models which couple tensors of rank one and two to the tensor models of \cite{Witten:2016iux, Gurau:2010ba, Gurau:2011aq}. We have argued that these describe the interaction of probes with a near extremal black hole, in particular these models exhibit maximal chaos. Mixed four point correlation functions of fields with different indices have leading order behavior which scales as $(\beta g_0)^2$ as opposed the four point functions of fields with identical indices which scales as $\beta g_0$. It would be interesting to understand this phenomena is greater detail.

We have uncovered the interesting feature that primaries made out of the probe fields develop the same scaling dimensions as those made out of black hole fields and we interpret this as the phenomena of the probe subsystem thermalizing with the bulk heat bath. In quantum many body systems thermalization is best understood in the light of Eigenstate Thermalization Hypothesis (ETH). Thus an obvious next step towards understanding this system better would be to study the validity of ETH in this model and we intend to investigate this in a future work. 
\bigskip

\vspace{1cm} \noindent {\bf Acknowledgements:} SM would like to thank Subhroneel Chakrabarti and Ashoke Sen for discussions. This work was conducted within the ILP LABEX (ANR-10-LABX-63) supported by French state funds managed by the ANR within the Investissements d'Avenir program (ANR-11-IDEX-0004-02) and supported partly by the CEFIPRA grant 5204-4. S.M. would like to thank Harish-Chandra Research Institite for hospitality, where part of the work was done.

\pagebreak
\appendix
\section{Conformal part of various four point functions} \label{fourdetails}
All the four point functions has the following form
\begin{align}
\sim \frac{K^l}{(1-K)^m} \mathcal{F}^0 \, . \label{fourpointform}
\end{align}
First we note that for $m=1$, the conformal piece is sum of residues at various simple poles\footnote{There is a double pole at $h=2$, but it cancells with the finite pieces coming from regulated $\mathcal{F}_2$ \cite{Maldacena:2016hyu}.} of $1/(1-K)$, where $k^l=1$. Thus the case for all $l$ are similar to $l=0$, which is the case for original SYK. Thus we refer the reader to \cite{Maldacena:2016hyu}.

For $m \neq 1$, the situation is different from that of original SYK. We start with $m=2$.
\begin{align}
\mathcal{F}' &= \frac{K^l}{(1-k)^2} \mathcal{F}_0 \, .
\end{align}
Let $P_{h=2}$ be the projector onto $h=2$ subspace and $P_{h \neq 2}$ its complement. Then 
\begin{align*}
\mathcal{F}' &= \frac{K^l}{(1-k)^2} \mathcal{F}_0 = \frac{K^l}{K-1} P_{h=2} \frac{1}{1-K} \mathcal{F}_0 + \frac{K^l}{K-1} P_{h \neq 2} \frac{1}{1-K} \mathcal{F}_0 \, .
\end{align*}
We are interested in the conformal piece. Since $K$ is diagonal in $|h\rangle$ basis, the conformal piece reads 
\begin{align}
\mathcal{F}'_{conf} &:= \frac{K^l}{K-1} P_{h \neq 2} \frac{1}{1-K} \mathcal{F}_0 \, .
\end{align}
Using $P_{h \neq 2} \frac{1}{1-K} \mathcal{F}_0  = \mathcal{F}_{conf}$, we have
\begin{align*}
\mathcal{F}'_{conf} &= \frac{K^l}{1-K} P_{h \neq 2} \mathcal{F}_{conf} = \sum_{h \neq 2} \frac{K^l}{1-K} \frac{|h\rangle \langle h | \mathcal{F}_{conf} \rangle }{\langle h| h \rangle} = \sum_{h \neq 2} \frac{k_c(h)^l}{1-k_c(h)} \frac{|h\rangle \langle h | \mathcal{F}_{conf} \rangle }{\langle h| h \rangle}
\end{align*}
As function of the $SL(2, \mathbb{R})$ invariant variable $\chi$ this reads
\begin{align}
\mathcal{F}'_{conf}(\chi) &=  \sum_{h \neq 2} \frac{k_c(h)^l}{1-k_c(h)} \frac{\Psi_h(\chi) \langle h | \mathcal{F}_{conf} \rangle }{\langle h| h \rangle}
\end{align}
The reader would note the similarity between the right hand side of the above equation and the corresponding one in \cite{Maldacena:2016hyu}. Only difference is that we have $\langle h| \mathcal{F}_{h \neq 2}\rangle$ where \cite{Maldacena:2016hyu} had $ \langle h| \mathcal{F}_{0}\rangle$. Since $\mathcal{F}_{h \neq 2}$ is well behaved, just like $\mathcal{F}_0$, the computation runs parallel to that of $\mathcal{F}_{h \neq 2}$ in \cite{Maldacena:2016hyu} and one gets
\begin{align}
-\sum_{m=1}^\infty \Res\left[ \frac{k_c(h)^l}{1 - k_c(h)} \frac{(2h-1)}{\pi \tan \frac{\pi h}{2}} \Psi_h(\chi) \langle h| \mathcal{F}_{conf} \rangle \right]_{h=h_m} &= \sum_{m=1}^\infty \frac{2h_m-1}{k_c'(h_m) \pi \tan \frac{\pi h_m}{2}} \Psi_{h_m}(\chi) \langle h_m | \mathcal{F}_{conf} \rangle \, , \label{res}
\end{align} 
where $k_c(h_m)=1$. Using 
\begin{align}
\mathcal{F}_{conf} (\chi) &=  \alpha_0 \sum_{m=1}^\infty \frac{(h_m -1/2)}{\pi k_c'(h_m) \tan \frac{\pi h_m}{2}} \Psi_{h_m}(\chi) \, , \label{m=1f}
\end{align}
and 
\begin{align}
\langle h| h' \rangle &= \frac{\pi^2 \tan \pi h }{2h-1} \delta(h - h') ~~~\text{for the continuum tower}~h=1/2+is \, ,
\end{align}
we have
\begin{align*}
\mathcal{F}'_{conf}(\chi) &= \frac{\alpha_0}{2} \sum_{m=1}^\infty \frac{(2h_m -1)}{k_c'(h_m)^2} \frac{\tan \pi h_m}{\tan^2 \frac{\pi h_m}{2}} \Psi_{h_m}(\chi) \, .
\end{align*}
In short time limit, i.e. $\chi \rightarrow 0$, this boils down to
\begin{align}
\mathcal{F}'_{conf}(\chi) &= \sum_{m=1}^\infty \tilde{c}_m^2 \chi^{h_m} ~~~~~
\text{where,}~~~\tilde{c}_m^2 = \frac{\alpha_0}{2} \frac{(2h_m -1)}{k'_c(h_m)^2} \frac{\tan \pi h_m}{\tan^2 \frac{\pi h_m}{2}}  \frac{\Gamma(h_m)^2}{\Gamma(2h_m)} \, . \label{1/(1-K)^2}
\end{align}
We have not taken\footnote{If one does not separate out $h \neq 2$ subspace from the beginning, as we have done, but computes the integral directly, then one has double poles instead of single poles and one gets a somewhat different expression for $\mathcal{F}'_{h \neq 2}$. We guess in such computation $\mathcal{F}'_2$ would be even more clumsy and its finite pieces would eventually combine with $\mathcal{F}'_{h \neq 2}$ to give back \refb{1/(1-K)^2}. It would be nice to check this.} the double pole at $h=2$. We hope this would cancel with finite contributions coming from regulated $\mathcal{F}'_{2}$, just as it does for SYK model. But we do not attempt to compute it here.

To move on to $m>2$, we note that we did not use the details of $\mathcal{F}_{h \neq 2}$ to get to \ref{res}. Thus if we denote the conformal part of $\frac{1}{(1-K)^m} \mathcal{F}_0$ as $\mathcal{F}_{h \neq 2}^{(m)}$, then we have the recursion relation
\begin{align}
\mathcal{F}_{conf}^{(m+1)} &= \sum_{m=1}^\infty \frac{2h_m-1}{k_c'(h_m) \pi \tan \frac{\pi h_m}{2}} \Psi_{h_m}(\chi) \langle h_m | \mathcal{F}^{(m)}_{conf} \rangle \, , \label{rec}
\end{align}
If 
\begin{align}
\mathcal{F}_{conf}^{(m)}(\chi) &= \sum_{k=1}^\infty (c_k^{(m)})^2 \Psi_{h_m}(\chi) \, ,
\end{align}
then we have the following recursion relation for $c^{(m)}$-s
\begin{align}
\left( c_k^{(m+1)} \right)^2 &= \left( c_k^{(m)} \right)^2 \frac{\pi \tan \pi h_k}{k_c'(h_k) \tan \frac{\pi h_k}{2}} \, .\label{recursion}
\end{align}
Using \ref{recursion} and \ref{m=1f}, we have
\begin{align}
\left( c_k^{(m)} \right)^2 &= \left( c_k^{(1)} \right)^2 \left[\frac{\pi \tan \pi h_k}{k_c'(h_k) \tan \frac{\pi h_k}{2}} \right]^{m-1} = 
\frac{ \alpha_0 (2h_k-1)\left[ \pi \tan \pi h_k \right]^{m-1}}{ 2\pi \left[ k_c'(h_k) \tan \frac{\pi h_k}{2} \right]^m} \, .
\end{align}

\section{Four point functions with maximal chaos} \label{maxchaos}
For any four point function with form 
\begin{align*}
\mathcal{F}_{m,n} &= \frac{K^m}{(1-K)^n} \mathcal{F}_0 \, ,
\end{align*}
the corresponding OTOC $F_{m,n}(t_1,t_2)$ will have maximal Lyapunov coefficient. We note that if we act $F_{m,n}$ by $K^R_{m,n} \equiv 1- (1-K_R)^n/K_R^m$, we get back $F_{m,n}$ except for a piece that can be ignored in large $t_1,t_2$ limit. Thus in this limit $F_{m,n}(t_1,t_2)$  is an eigenfunction of the operator $K^R_{m,n}$ with eigenvalue $1$. Now the operator $K^R_{m,n}$ clearly commutes with $K_R$ and thus has same eigenfunctions, labelled by $h$ with eigenvalues given by
\begin{align}
k_{m,n}(h)&= 1-(1-k_R(h))^n/k_R(h)^m \, .
\end{align} 
It is clear that $k_{m,n}(h)=1\Rightarrow k_R(h)=1$. We have previously mentioned (see \ref{chaosection}) that $k_R(h)=1$ is satisfied for $h=-1$ with the eigenfunction $e_{-1}(t_1,t_2)$ given in \ref{kreigen}. This essentially implies that in long time limit $F_{m,n}=e_{-1}$ and therefore has maximal Lyapunov coefficient $\lambda_L=2\pi/\beta$.


\bibliographystyle{/Users/Halmagyi/Dropbox/utphys} 
\bibliography{/Users/Halmagyi/Dropbox/myrefs}

\providecommand{\href}[2]{#2}\begingroup\raggedright\begin{thebibliography}{10}

\bibitem{K2}
A.~Kitaev, ``{KITP strings seminar and Entanglement 2015 program},''.
  http://online.kitp.ucsb.edu/online/entangled15/.

\bibitem{Maldacena:2016hyu}
J.~Maldacena and D.~Stanford, ``{Remarks on the Sachdev-Ye-Kitaev model},''
  {\em Phys. Rev.} {\bf D94} (2016), no.~10, 106002,
\href{http://arXiv.org/abs/1604.07818}{{\tt 1604.07818}}.

\bibitem{Maldacena:2016upp}
J.~Maldacena, D.~Stanford, and Z.~Yang, ``{Conformal symmetry and its breaking
  in two dimensional Nearly Anti-de-Sitter space},'' {\em PTEP} {\bf 2016}
  (2016), no.~12, 12C104,
\href{http://arXiv.org/abs/1606.01857}{{\tt 1606.01857}}.

\bibitem{Garcia-Garcia:2017pzl}
A.~M. Garcia-Garcia and J.~J.~M. Verbaarschot, ``{Analytical Spectral Density
  of the Sachdev-Ye-Kitaev Model at finite N},'' {\em Phys. Rev.} {\bf D96}
  (2017), no.~6, 066012,
\href{http://arXiv.org/abs/1701.06593}{{\tt 1701.06593}}.

\bibitem{Sonner:2017hxc}
J.~Sonner and M.~Vielma, ``{Eigenstate thermalization in the Sachdev-Ye-Kitaev
  model},''
\href{http://arXiv.org/abs/1707.08013}{{\tt 1707.08013}}.

\bibitem{Fu:2016vas}
W.~Fu, D.~Gaiotto, J.~Maldacena, and S.~Sachdev, ``{Supersymmetric
  Sachdev-Ye-Kitaev models},'' {\em Phys. Rev.} {\bf D95} (2017), no.~2,
  026009, \href{http://arXiv.org/abs/1610.08917}{{\tt 1610.08917}}.
[Addendum: Phys. Rev.D95,no.6,069904(2017)].

\bibitem{Hunter-Jones:2017raw}
N.~Hunter-Jones, J.~Liu, and Y.~Zhou, ``{On thermalization in the SYK and
  supersymmetric SYK models},''
\href{http://arXiv.org/abs/1710.03012}{{\tt 1710.03012}}.

\bibitem{Berkooz:2016cvq}
M.~Berkooz, P.~Narayan, M.~Rozali, and J.~Simón, ``{Higher Dimensional
  Generalizations of the SYK Model},'' {\em JHEP} {\bf 01} (2017) 138,
\href{http://arXiv.org/abs/1610.02422}{{\tt 1610.02422}}.

\bibitem{Turiaci:2017zwd}
G.~Turiaci and H.~Verlinde, ``{Towards a 2d QFT Analog of the SYK Model},''
  {\em JHEP} {\bf 10} (2017) 167,
\href{http://arXiv.org/abs/1701.00528}{{\tt 1701.00528}}.

\bibitem{Gross:2016kjj}
D.~J. Gross and V.~Rosenhaus, ``{A Generalization of Sachdev-Ye-Kitaev},'' {\em
  JHEP} {\bf 02} (2017) 093,
\href{http://arXiv.org/abs/1610.01569}{{\tt 1610.01569}}.

\bibitem{Polchinski:2016xgd}
J.~Polchinski and V.~Rosenhaus, ``{The Spectrum in the Sachdev-Ye-Kitaev
  Model},'' {\em JHEP} {\bf 04} (2016) 001,
\href{http://arXiv.org/abs/1601.06768}{{\tt 1601.06768}}.

\bibitem{Gross:2017hcz}
D.~J. Gross and V.~Rosenhaus, ``{The Bulk Dual of SYK: Cubic Couplings},'' {\em
  JHEP} {\bf 05} (2017) 092,
\href{http://arXiv.org/abs/1702.08016}{{\tt 1702.08016}}.

\bibitem{Gross:2017aos}
D.~J. Gross and V.~Rosenhaus, ``{All point correlation functions in SYK},''
\href{http://arXiv.org/abs/1710.08113}{{\tt 1710.08113}}.

\bibitem{Das:2017pif}
S.~R. Das, A.~Jevicki, and K.~Suzuki, ``{Three Dimensional View of the SYK/AdS
  Duality},'' {\em JHEP} {\bf 09} (2017) 017,
\href{http://arXiv.org/abs/1704.07208}{{\tt 1704.07208}}.

\bibitem{Maldacena:2015waa}
J.~Maldacena, S.~H. Shenker, and D.~Stanford, ``{A bound on chaos},'' {\em
  JHEP} {\bf 08} (2016) 106,
\href{http://arXiv.org/abs/1503.01409}{{\tt 1503.01409}}.

\bibitem{Witten:2016iux}
E.~Witten, ``{An SYK-Like Model Without Disorder},''
\href{http://arXiv.org/abs/1610.09758}{{\tt 1610.09758}}.

\bibitem{Gurau:2010ba}
R.~Gurau, ``{The 1/N expansion of colored tensor models},'' {\em Annales Henri
  Poincare} {\bf 12} (2011) 829--847,
\href{http://arXiv.org/abs/1011.2726}{{\tt 1011.2726}}.

\bibitem{Gurau:2011aq}
R.~Gurau and V.~Rivasseau, ``{The 1/N expansion of colored tensor models in
  arbitrary dimension},'' {\em EPL} {\bf 95} (2011), no.~5, 50004,
\href{http://arXiv.org/abs/1101.4182}{{\tt 1101.4182}}.

\bibitem{Klebanov:2016xxf}
I.~R. Klebanov and G.~Tarnopolsky, ``{Uncolored random tensors, melon diagrams,
  and the Sachdev-Ye-Kitaev models},'' {\em Phys. Rev.} {\bf D95} (2017),
  no.~4, 046004,
\href{http://arXiv.org/abs/1611.08915}{{\tt 1611.08915}}.

\bibitem{Peng:2016mxj}
C.~Peng, M.~Spradlin, and A.~Volovich, ``{A Supersymmetric SYK-like Tensor
  Model},'' {\em JHEP} {\bf 05} (2017) 062,
\href{http://arXiv.org/abs/1612.03851}{{\tt 1612.03851}}.

\bibitem{Krishnan:2016bvg}
C.~Krishnan, S.~Sanyal, and P.~N. Bala~Subramanian, ``{Quantum Chaos and
  Holographic Tensor Models},'' {\em JHEP} {\bf 03} (2017) 056,
\href{http://arXiv.org/abs/1612.06330}{{\tt 1612.06330}}.

\bibitem{Choudhury:2017tax}
S.~Choudhury, A.~Dey, I.~Halder, L.~Janagal, S.~Minwalla, and R.~Poojary,
  ``{Notes on Melonic $O(N)^{q-1}$ Tensor Models},''
\href{http://arXiv.org/abs/1707.09352}{{\tt 1707.09352}}.

\bibitem{Bulycheva:2017ilt}
K.~Bulycheva, I.~R. Klebanov, A.~Milekhin, and G.~Tarnopolsky, ``{Spectra of
  Operators in Large $N$ Tensor Models},''
\href{http://arXiv.org/abs/1707.09347}{{\tt 1707.09347}}.

\bibitem{Peng:2017kro}
C.~Peng, ``{Vector models and generalized SYK models},'' {\em JHEP} {\bf 05}
  (2017) 129,
\href{http://arXiv.org/abs/1704.04223}{{\tt 1704.04223}}.

\bibitem{Yoon:2017nig}
J.~Yoon, ``{SYK Models and SYK-like Tensor Models with Global Symmetry},'' {\em
  JHEP} {\bf 10} (2017) 183,
\href{http://arXiv.org/abs/1707.01740}{{\tt 1707.01740}}.

\bibitem{Iizuka:2008hg}
N.~Iizuka and J.~Polchinski, ``{A Matrix Model for Black Hole
  Thermalization},'' {\em JHEP} {\bf 0810} (2008) 028,
\href{http://arXiv.org/abs/0801.3657}{{\tt 0801.3657}}.

\bibitem{Iizuka:2008eb}
N.~Iizuka, T.~Okuda, and J.~Polchinski, ``{Matrix Models for the Black Hole
  Information Paradox},'' {\em JHEP} {\bf 1002} (2010) 073,
\href{http://arXiv.org/abs/0808.0530}{{\tt 0808.0530}}.

\bibitem{Michel:2016kwn}
B.~Michel, J.~Polchinski, V.~Rosenhaus, and S.~J. Suh, ``{Four-point function
  in the IOP matrix model},'' {\em JHEP} {\bf 05} (2016) 048,
\href{http://arXiv.org/abs/1602.06422}{{\tt 1602.06422}}.

\bibitem{Bonzom:2012hw}
V.~Bonzom, R.~Gurau, and V.~Rivasseau, ``{Random tensor models in the large N
  limit: Uncoloring the colored tensor models},'' {\em Phys. Rev.} {\bf D85}
  (2012) 084037,
\href{http://arXiv.org/abs/1202.3637}{{\tt 1202.3637}}.

\bibitem{Bagrets:2017pwq}
D.~Bagrets, A.~Altland, and A.~Kamenev, ``{Power-law out of time order
  correlation functions in the SYK model},'' {\em Nucl. Phys.} {\bf B921}
  (2017) 727--752,
\href{http://arXiv.org/abs/1702.08902}{{\tt 1702.08902}}.

\bibitem{Dartois:2017xoe}
S.~Dartois, H.~Erbin, and S.~Mondal, ``{Conformality of $1/N$ corrections in
  SYK-like models},''
\href{http://arXiv.org/abs/1706.00412}{{\tt 1706.00412}}.

\end{thebibliography}\endgroup

\end{document}